\documentclass{nature}

\usepackage{graphicx}
\usepackage{verbatim}
\usepackage{amsfonts} 
\usepackage{amsmath}
\usepackage{mathbbol}
\usepackage{amssymb}             

\DeclareSymbolFontAlphabet{\amsmathbb}{AMSb}%

\newcommand{\be}{\begin{equation}}
\newcommand{\ee}{\end{equation}}

\makeatletter
\let\saved@includegraphics\includegraphics
\AtBeginDocument{\let\includegraphics\saved@includegraphics}
\renewenvironment*{figure}{\@float{figure}}{\end@float}
\makeatother

\begin{document}






\bibliographystyle{naturemag}

\title{Casimir force inadequacy in explaining a strong attractive force in a micrometer-sized narrow-gap re-entrant cavity}



\author{Giuseppe Bimonte$^{1,2}$}

\maketitle

\begin{affiliations}
\item Dipartimento di Fisica E. Pancini, Universit\'a di Napoli Federico II,
Complesso Universitario di Monte S. Angelo,
Via Cinthia, I-80126 Napoli, Italy
\item INFN Sezione di Napoli, I-80126 Napoli, Italy
\end{affiliations}



Pate et al. \cite{pate} investigated a macroscopic opto-mechanical system with a narrow-gap re-entrant cavity coupled to a SiN membrane resonator coated with Au or Nb. They observed a significant increase in the membrane's effective spring constant $k_{\rm eff}$ for sub-2-micron gaps $x$. This increase scales roughly with $x^{-4}$, suggesting an attractive force pulling the membrane towards the  re-entrant  Al post, with an $x^{-3}$ dependence. Attributing this force solely to the thermal Casimir effect is challenged by our detailed calculations (presented below). These calculations reveal that the Casimir force, at the investigated gap sizes, is orders of magnitude weaker than the observed force. This significant discrepancy necessitates an alternative explanation for the observed attraction.

\section*{Computation of the Casimir spring}

The  geometry of the re-entrant  the cavity  is displayed in Fig. 3 of the Supplementary material to the paper \cite{pate}. 
The Casimir force $F_C(x)$ between the Al post and the membrane can be estimated using  the standard Proximity Force Approximation (PFA) \cite{parsegian,bordag}, which consists in decomposing the surfaces of the two bodies  into pairs of small and parallel patches, and then adding up the Casimir forces for all pairs of patches.  The Proximity Force Approximation (PFA) remains a popular tool for interpreting Casimir force measurements due to its simplicity and effectiveness for objects in close proximity. This is particularly true for the narrow-gap cavity used by Pate et al. \cite{pate}, where the gap size $x$ is significantly smaller than the post's cap radius $r_0$ by a factor exceeding 50.
It's important to remember that the experiment is primarily concerned with the Casimir spring constant $k_{C}$ rather than the absolute force. The spring constant represents the rate of change of the Casimir force $F_{\rm C}(x)$ with respect to the gap size $x$:
\begin{equation}
k_C=F'_C(x)\;.
\end{equation} 
Using the PFA, one finds:
\begin{equation}
k_C=\pi r_0^2F'_{\rm PP}(x) + \frac{2 \pi (r_1-r_0)}{h}\left[r_1 F_{\rm PP}(x+h) -r_0 F_{\rm PP}(x)  \right]+ \frac{2 \pi (r_1-r_0)^2}{h^2}\left[E_{\rm PP}(x+h)-E_{\rm PP}(x) \right]\,, \label{PFA}
\end{equation}
where the geometric parameters $r_0,r_1, h$ are defined as in Fig. 3 of the Supplementary material to \cite{pate}. In the above formula, $E_{\rm PP}(a)$ represents the Casimir energy  per unit-area between two (infinite) plane-parallel slabs separated by a gap of width $a$, while $F_{\rm PP}(a)=-E'_{\rm PP}(a)$ is the corresponding Casimir force per unit area (negative forces represent attraction).  
The first term in the equation accounts for the contribution of the post's top flat surface, while the remaining terms  represent the contribution of the post's sidewalls.  According to Lifshitz formula \cite{lifshitz} $E_{\rm PP}(a)$ has the expression:
\be
E_{\rm PP}(a)=\frac{k_B T}{2\,\pi}\sum_{l=0}^{\infty}\;\!\!' \int_0^{\infty}\,dk_{\perp} k_{\perp}  
\sum_{\alpha={\rm TE, TM}} \log \left[ 1- r^{(1)}_{\alpha}(i \xi_l, k_{\perp}) r^{(2)}_{\alpha}(i \xi_l, k_{\perp}) \,e^{-2 a q_l} \right]\;,\label{lifs}
\ee  
where $k_B$ is Boltzmann constant, $T$ is the temperature,  $k_{\perp}$ is the in-plane momentum, the prime in the sum indicates that the  $l=0$ term is taken with weight one-half, $\xi_l= 2 \pi l k_B T/\hbar$ are the imaginary Matsubara frequencies, $q_l=\sqrt{\xi_l^2/c^2+k_{\perp}^2}$,  the index $\alpha={\rm TE, TM}$ labels the two independent states of polarization of the electromagnetic field, i.e. transverse magnetic ($\rm TM$) and transverse electric  ($\rm TE$), and $r^{(k)}_{\alpha}(i \xi_l, k_{\perp}) $ denote the Fresnel reflection coefficients of the $k$-th slab for polarization $\alpha$. 
We modeled the aluminum post (slab 1) as infinitely thick, while the membrane (slab 2) is a composite of a 500 nm SiN substrate and a 300 nm metallic (Au or Nb) coating. The force  $F_{\rm PP}(a)$ and  and its derivative $F'_{\rm PP}(a)$ relevant to the Casimir force are obtained from Eq. (~\ref{lifs}). The experimental setup in Pate et al. \cite{pate} allows a simplification. The gap size $x$ is much smaller than characteristic dimensions of the post ($r_0$, $r_1$, and $h$), and moreover the post is thin ($h \gg r_0,r_1$). This implies that the post's lateral surface has a negligible contribution to the Casimir spring constant $k_{\rm C}$ compared to its top face. Therefore, in Eq. (~\ref{PFA}), only the first term is significant. Consequently, the Casimir spring constant can be approximated by a simpler formula:
\begin{equation}
k_C=\pi r_0^2F'_{\rm PP}(x) \,. \label{PFA2}
\end{equation}
Another key simplification emerges from the properties of relevant Casimir force contributors. Lifshitz theory (Eq. (~\ref{lifs})) indicates that crucial Matsubara modes have imaginary frequencies near $\omega_c = c/(2x)$, determined by the gap size $x$. For the experiment's gap range (0.59 $\mu$m to 3.3 $\mu$m), the penetration depth $\delta$ of these modes in gold (Au), niobium (Nb), and aluminum (Al) is limited to tens of nanometers. Since this depth is significantly smaller than the metallic coating thickness (300 nm) on the SiN membrane, the membrane behaves essentially like an infinitely thick slab of either Au or Nb for Casimir force calculations. This allows us to model both the post and the membrane as infinitely thick planar slabs (Al and Au/Nb, respectively) when evaluating Eq. (~\ref{PFA2}). Consequently, the following well-known expressions for Fresnel coefficients can be employed:
\be
r^{(k)}_{\rm TE}(i \xi_l, k_{\perp}) =\frac{q_l-s^{(k)}_l}{q_l+s^{(k)}_l}\;,\label{TE}
\ee
\be
r^{(k)}_{\rm TM}(i \xi_l, k_{\perp}) =\frac{\epsilon^{(k)}_l q_l-s^{(k)}_l}{\epsilon^{(k)}_l\,q_l+s^{(k)}_l}\;,\label{TM}
\ee
where $s^{(k)}_l=\sqrt{\epsilon^{(k)}_l \xi_l^2/c^2+k_{\perp}^2}$, and $\epsilon^{(k)}_l \equiv \epsilon^{(k)}(i \xi_l)$ is the permittivity of the material constituting the  slab. In the range of frequencies that are relevant to the Casimir force, the optical properties of  Au, Nb and Al
can be described by a simple Drude model
 \begin{equation}
\epsilon(i\xi) =1+  \frac{\Omega^2}{\xi(\xi+\gamma)} \, ,
\end{equation}
where $\Omega$ is the plasma frequency and $\gamma$ is the relaxation frequency
 The parameters are provided in Tab.~\ref{tab:1} in units of eV/$\hbar$. 
\begin{figure}
\includegraphics[width=0.6\textwidth]{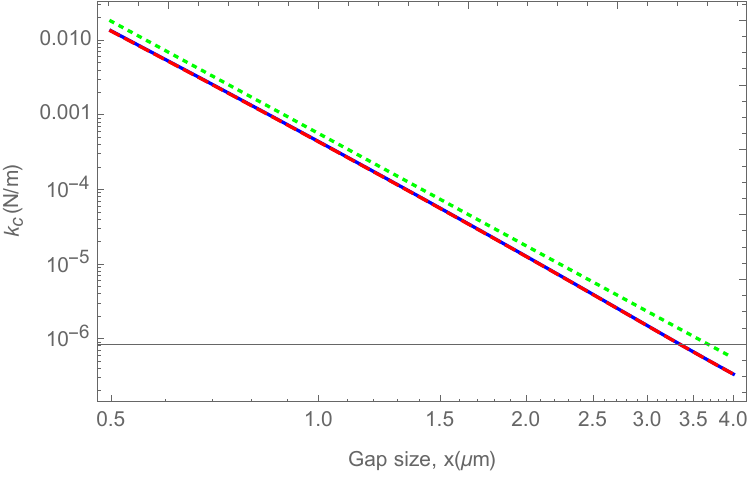} 
\caption{\label{fig:1}
Casimir spring constant $k_{\rm C}=F'_{\rm C}(x)$ as a function of gap size, $x$. The spring constant is shown for an Au-coated membrane (blue solid line) a Nb-coated membrane (red dashed line), both computed for room temperature $T=300$ K.  The green dotted line shows the Casimir spring constant $k^{\rm (pc)}_{\rm C}$ in the limit of a perfectly conducting cavity at $T=0$ (see Eq. (\ref{pc})).}
\end{figure}
In Fig. 1 we show a plot of the Casimir spring $k_{\rm C}$ (in N/m) for the gold coated membrane (blue solid line) and for a Nb coated membrane (red dashed line), versus the gap size $x$ (in $\mu$m), computed using Eq. (\ref{PFA2}) for room temperature $T=300$ K.  The green dotted curve in Fig. 1 shows the PFA value $k^{\rm (pc)}_{\rm C}$ of  the spring constant  Eq. (\ref{PFA2}) in the limit of a perfectly conducting cavity at $T=0$:
\be
k^{(\rm pc)}_{\rm C}=  \frac{\pi^3 \hbar c\,r_0^2}{60 \; x^5}\;.\label{pc}
\ee 
 \section*{Discussion}
Our calculations (Fig. 1) reveal that the Casimir spring constant, $k_{\rm C}$, is significantly weaker - orders of magnitude lower - than the fundamental spring constant, $k_{\rm S}$, of the membranes (572 N/m for Au and 949 N/m for Nb, as shown by the flat orange and gray lines in Fig. 2 of Pate et al. \cite{pate}). This vast discrepancy eliminates the need for more computationally expensive, exact methods like those described in \cite{emigMSE} to refine our calculations.  Since $k_{\rm C}$ is so much weaker than $k_{\rm S}$, the Casimir force cannot be the primary explanation for the substantial increase in the effective spring constant, $k_{\rm eff}$, observed by Pate et al. for gaps below 2 microns.
Since the Casimir force cannot explain the observed attraction, alternative explanations must be explored.  Assuming the experiment is accurate, a possible explanation lies in electrostatic interactions between the aluminum post and the membrane.  Previous research  \cite{speake} suggests that variations in surface potential across these surfaces can lead to a significant electrostatic force. This electrostatic force might exhibit a similar distance dependence (scaling with the gap size) as the Casimir force, but with a much larger magnitude. This aligns with observations from Sushkov et al. \cite{Sushkov:2011}, where electrostatic forces due to surface potential variations were proposed to explain forces ten times stronger than the Casimir force in a torsional balance experiment with gold coated surfaces. To investigate the possibility of electrostatic effects, a Kelvin probe measurement \cite{behunin} could be a valuable tool. By mapping the electrostatic potential across the surfaces used in Pate et al.'s experiment \cite{pate}, this technique could reveal the presence or absence of significant potential variations.



\subsection{Data availability}
The data that support the findings of this study are available from the corresponding author, upon reasonable request.



\subsection{References}

\bibliographystyle{naturemag}

\bibliography{spring.bib}


\begin{addendum}
 \item Discussions with A. Cassinese are acknowledged.
 \item[Competing Interests] The author declares that he has no
competing financial interests.
 \item[Correspondence] Correspondence and requests for materials
should be addressed to G.B..~(email: bimonte@na.infn.it).
\end{addendum}


\begin{table}
  \centering
\caption{\label{tab:1} Drude parameters for gold, niobium and aluminum.}
\medskip
\begin{tabular}{ll}
\hline
\hline
Parameters for Al & value [eV/$\hbar$] \\
\hline
$\Omega_{\rm Al}$  & 13\\
$\gamma_{\rm Al}$  & 0.1\\
\hline
Parameters for Au & \\
\hline
$\Omega_{\rm Au}$ & 9.0\\
$\gamma_{\rm Au}$ & 0.035\\
\hline
Parameters for Nb & \\
\hline
$\Omega_{\rm Nb}$ & 9.9\\
$\gamma_{\rm Nb}$ & 0.2\\
\hline
\hline
\end{tabular}
\end{table}

\end{document}